\documentclass[12pt]{article}
\usepackage{a4wide,graphicx,amsmath,amssymb,hyperref}
\usepackage[numbers,sort&compress]{natbib}
\usepackage{hypernat}
\usepackage{subfigure}
\usepackage{multirow}
\usepackage{color}
\usepackage{ulem}

\renewcommand{\vec}[1]{\boldsymbol{#1}}

\def\beq{\begin{equation}}
\def\eeq{\end{equation}}
\def\bea{\begin{eqnarray}}
\def\eea{\end{eqnarray}}

\newcommand{\newc}{\newcommand}

\newc{\scale}{0.75}
\newc{\twoscale}{0.74}
\newc{\twoFeynScale}{0.45}

\newc{\atxt}[1]{{\color{blue}\em #1}}
\newc{\jtxt}[1]{{\color{red}\em #1}}
\newc{\stxt}[1]{{\color{magenta}\em #1}}
\newc{\cmt}[1]{{\color{green}\bf\em #1}}

\newc{\Quote}[1]{``#1''}
\newc{\cf}{\textit{cf}.~}
\newc{\ie}{i.e.~}
\newc{\wrt}{w.r.t.~}
\newc{\eg}{e.g.~}

\newc{\ifb}{\textrm{fb}^{-1}}

\newc{\fb}{{\rm fb}}
\newc{\pb}{{\rm pb}}
\newc{\nb}{{\rm nb}}
\newc{\mb}{{\rm mb}}

\newc{\sigeff}{\sigma_{\rm eff}}
\newc{\sigDPS}{\sigma_{\rm DPS}}
\newc{\sigSPS}{\sigma_{\rm SPS}}

\newc{\muf}{\mu_{F}}
\newc{\mur}{\mu_{R}}

\newc{\mstwlo}{\texttt{MSTW 08 LO}~}
\newc{\mstwnl}{\texttt{MSTW 08 NLO}~}

\newc{\Jpsi}{J/\psi}
\newc{\jpsi}{\Jpsi}
\newc{\DDY}{\mathrm{DDY}}
\newc{\DY}{\mathrm{DY}}
\newc{\mup}{\mu^+}
\newc{\mum}{\mu^-}
\newc{\dmu}{\mup\mum}
\newc{\gams}{\gamma^*}
\newc{\Wpm}{W^{\pm}}

\newc{\mmumu}{m_{\dmu}}
\newc{\pt}{p_{T}}
\newc{\vpt}{\vec{\pt}}
\newc{\mpt}{\langle p_{T}\rangle}
\newc{\dR}{\Delta R}
\newc{\dphi}{\Delta \phi}
\newc{\deta}{\Delta \eta}
\newc{\rap}{y}
\newc{\drap}{\Delta\rap}
\newc{\drapmin}{|\Delta\rap|_{\rm min}}
\newc{\shat}{\hat{s}}
\newc{\rtshat}{\sqrt{\shat}}

\newc{\pta}{p_{T\mup_1}}
\newc{\ptb}{p_{T\mum_1}}
\newc{\ptc}{p_{T\mup_2}}
\newc{\ptd}{p_{T\mum_2}}

\newc{\br}{\mathcal{BR}}
\newc{\ord}{\mathcal{O}}

\newc{\herwig}{\texttt{Herwig++}~}
\newc{\herwigv}{\texttt{Herwig++ v2.4.2}~}
\newc{\madgraph}{\texttt{MADGRAPH}~}
\newc{\madgraphv}{\texttt{MADGRAPH v5.1.2.4}~}
\newc{\mcfm}{\texttt{MCFM}~}

\begin{document}
  \titlepage
  \begin{flushright}
    Cavendish-HEP-2011/15 \\
    DAMTP-2011-56\\
    TTK-11-27
  \end{flushright}
  \vspace*{0.5cm}
  \begin{center}
    {\Large \bf Prospects for observation of double parton scattering with four--muon
      final states at LHCb}\\
    \vspace*{1cm}
    \textsc{C.H.~Kom$^{a,b}$, A.~Kulesza$^c$ and W.J.~Stirling$^a$}\\
    \vspace*{0.5cm}
{\it
         $^a$ Cavendish Laboratory, University of Cambridge, CB3 0HE, UK\\
         $^b$ Department of Applied Mathematics and Theoretical Physics, University of Cambridge, CB3 0WA, UK\\
         $^c$ Institute for Theoretical Particle Physics and Cosmology, RWTH Aachen University, D-52056 Aachen, Germany
}
  \end{center}
  \vspace*{0.5cm}
  \begin{abstract}
We study the prospects for observing double parton scattering through
four--muon final states, forming two opposite--sign muon pairs, in the
LHCb experiment in $pp$ collisions at 14 TeV centre of mass energy.
We consider two special cases, namely double Drell--Yan and
$\jpsi$--pair production.  The kinematic properties and prospects for
observing these processes are discussed.  We find that the production
rate depends strongly on the origin of the four muons, while many
kinematic properties can be used to help identify the presence of
double parton scattering events.
  \end{abstract}


\section{Introduction}\label{sec:intro}

The subject of multiple parton interactions has been enjoying a
renewed interest over the last few years, driven by the need to
understand the full spectrum of hadronic activity at the LHC.  The
large collision energy available at the LHC increases the probability
of scattering of more than just one parton from each hadron. In
particular, multiple scattering events with final state particles
carrying relatively high transverse momentum should occur much more
commonly than at the previously operating hadron colliders. Such
observations provide valuable information on the structure of the
proton and the parton--parton correlations within.  Indeed, processes
involving two hard scatterings in one hadron--hadron collision, the
so--called double parton scattering (DPS), have been studied in the
4--jet and the $\gamma$+3--jet channels by the AFS~\cite{AFS},
UA2~\cite{UA2}, CDF~\cite{CDF} and D0~\cite{D0} experiments, and the
presence of DPS is now well established.

In order to measure double parton scattering, one needs to consider
processes for which the single parton scattering (SPS) background is
low or possesses characteristics that allow for the signal to be
extracted. To date, phenomenological studies have focused on
observation of DPS in hadronic processes such as four
jets~\cite{Paver:1982yp,Humpert:1983pw,Humpert:1984ay,Ametller:1985tp,Mangano:1988sq,Berger:2009cm,Domdey:2009bg,Blok:2010ge,
  Blok:2011bu}, $b\bar{b}$
pairs~\cite{DelFabbro:2002pw,Cattaruzza:2005nu}, $b\bar{b}$ pair and
two jets~\cite{Berger:2009cm}, Higgs and a gauge
boson~\cite{DelFabbro:1999tf,Hussein:2006xr}, production of either
photons~\cite{Drees:1996rw,Eboli:1997sv},
$W$~\cite{Godbole:1989ti,Maina:2009vx,Maina:2009sj} or $Z$
bosons~\cite{Maina:2009sj} in association with jets, or purely lepton
events such as double Drell--Yan~\cite{Goebel:1979mi,Halzen:1986ue}
and same--sign W--pair
production~\cite{Kulesza:1999zh,Cattaruzza:2005nu, Maina:2009sj,
  Gaunt:2010pi}.

In the environment of hadronic collisions, purely leptonic signatures
offer a clean probe of the underlying scattering mechanisms.  A
generic property of DPS processes is that they are expected to peak
strongly in the low transverse momentum ($\pt$), low $Q^2$ phase space
region, which is difficult to access with a typical central detector.
In final states involving jets, tagging and identifying low $\pt$ jets
associated with the DPS signal process is challenging due to the
overwhelming QCD background.  However multi--lepton, especially muon,
final states are much cleaner, making triggering on multi lepton final
states with low $\pt$ possible.  In this paper we take advantage of
the excellent low $\pt$ muon acceptance of the LHCb detector, which
can go down to $\sim$ 1 GeV, and focus on four--muon final states that
form two opposite--sign (OS) muon pairs in this particular experiment.
This should maximize the number of DPS events observed.

Depending on the analysis cuts, in particular on the invariant mass of
the OS muon pairs, the DPS signal could be dominated by double
Drell--Yan (DDY) production or production of a quarkonium (\eg $\jpsi$
or $\Upsilon$) pair, which then decay into four leptons.  The
possibility of observing DPS in pair--production of quarkonia has been
recently investigated in~\cite{Kom:2011bd, Baranov:2011ch,
  Novoselov:2011ff}.  In particular, in Ref.~\cite{Kom:2011bd} we have
argued that the recent measurement of double $\jpsi$ production by the
LHCb experiment~\cite{LHCb-CONFNote_DJpsi} could already indicate a
significant contribution from DPS.

While the production of double $\jpsi$ pairs benefits from a large
cross section, the subprocess of DDY, \ie {\it single} DY production,
is theoretically well understood.  For this reason, double Drell--Yan
production might be considered a standard--candle DPS process and
merits a more detailed investigation.  Moreover, the partons involved
in these processes are different.  DPS double $\jpsi$ production is
expected to be dominantly produced by four gluons, whereas for DDY,
the initial states at leading order (LO) are two quark--anti--quark
pairs.  Hence the correlations being probed are different.  In this
paper, we shall primarily focus on the prospect for observing DPS
through the DDY channel.  We also comment on some characteristic
differences between double $\jpsi$ and DDY production.

The paper is structured as follows. In Section \ref{sec:process} we
discuss the DPS signal and SPS background to DDY production at the
LHC. Our method of performing calculations and cuts chosen to select
the signal appropriate to the LHCb experiment are presented in
Section~\ref{sec:eventSim}, where we also give the total rates for the
processes of interest. In Section \ref{sec:kinematics} we address the
differential distributions for the signal and background to the DDY
process, followed by a discussion of differences between the DDY and
double $\jpsi$ distributions in Section \ref{sec:kinematics_jpsi},
before concluding in Section \ref{sec:conclusion}.


\section{Double Drell--Yan at LHCb}\label{sec:process}

\subsection{DPS signal}\label{subsec:process_DPS}

\begin{figure*}[!t]
  \begin{center}
    \scalebox{\scale}{
      \includegraphics{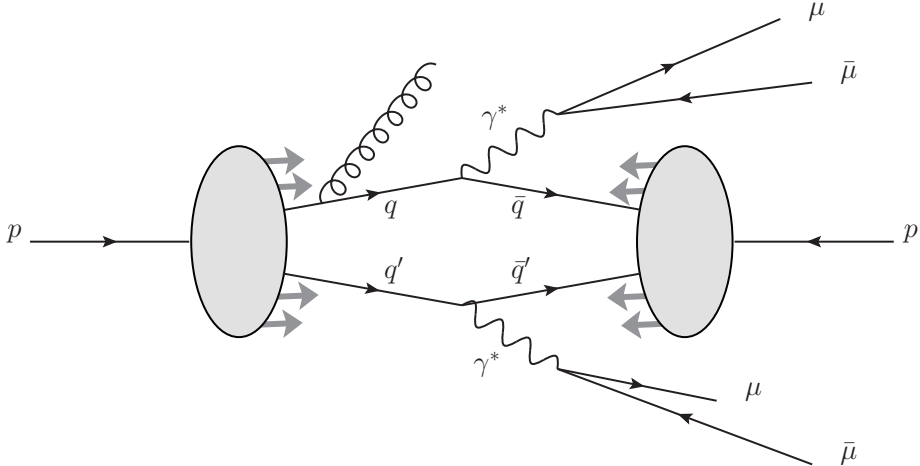}
    }
    \caption{Example Feynman diagram for the DPS double Drell--Yan process.}
    \label{fig:DPSFeynDiag}
  \end{center}
\end{figure*}

The DPS signal process leading to the two OS muon pair final state is
assumed to come from two independent hard scatterings of the DY type.
An example Feynman diagram is shown in Fig.~\ref{fig:DPSFeynDiag}.  If
we further assume that the longitudinal and the transverse components
of the generalized double parton distribution functions can be
factorised, and that there are no longitudinal momentum correlations
between the partons in the same hadron, the DPS DDY cross section
factorizes into a product of two SPS {\it single} DY cross
sections. More precisely, the differential DPS cross section
$d\sigDPS^{\DDY}$ is written as
\begin{equation}
  \label{eq:DPS}
  d \sigDPS^{\DDY} = 
  \frac{d \sigSPS^{\DY} \, d \sigSPS^{\DY}}{2 \sigeff},
\end{equation}
where
\begin{equation}
  \label{eq:SPS}
  d \sigSPS^{\DY} = \sum_{a,b} f_a (x_a,\mu_F) f_b (x_b,\mu_F)\, d \hat{\sigma}_{\rm SPS}^{\DY} \, d x_a dx_b \,,
\end{equation}
and
\begin{equation}
  \label{eq:partSPS}
  d \hat{\sigma}_{\rm SPS}^{\DY} = \sum_{a,b} \, \frac{1}{2 \hat{s}} \,
  \overline{| M_{a b \to \gamma^* \to l^+ l^- +X} |^2} \, d{\rm PS}_{l^+
    l^- +X }\, 
\end{equation} 
correspond to the hadronic and partonic single DY process
respectively.  In the above expressions, $\hat{s}$ is the partonic
centre--of--mass energy squared, $f_i$ and $x_i$ ($i=a,b$) are the
parton distribution functions (PDFs) and longitudinal momentum
fraction for parton with flavour $i$, and $\muf$ is the factorization
scale.  The single DY matrix element and phase space element are
denoted by $M$ and $d{\rm PS}$ respectively.  The factor $\sigeff$
captures all the information on the transverse structure of the
proton.  It is likely to be process and energy dependent, and is the
main quantity that a DPS experiment will aim to extract.  However, for
concreteness, we shall use $\sigeff=14.5$ $\mb$, a value measured in
the $\gamma$+3--jet study by the CDF experiment~\cite{CDF}.\footnote{
  A similar value, 15.1 $\mb$, was obtained by the corresponding D0
  experiment\cite{D0}.}

Recently, questions have been raised~\cite{Diehl:2011tt,Gaunt:2011xd}
about to what extent this intuitive picture can be derived in
perturbative QCD.  In particular, the assumption of factorization
between the longitudinal and transverse components of double parton
distributions appears not to be strictly consistent with QCD. It leads
to an appearance of the spurious \Quote{single--feed} term in the
evolution equation for double parton distributions, describing
correlations in longitudinal momentum fractions $x_i$. However, in the
current analysis we consider final states produced through DPS with
relatively low invariant masses, meaning the incoming partons have on
average low longitudinal momentum. Since at low $x$ the parton--parton
correlations are expected to be negligible, using single parton
distributions $f_i(x_i,\mu_F)$ is likely to be a good approximation.

\subsection{SPS background}\label{subsec:process_SPS}

\begin{figure*}[!t]
  \begin{center}
    \subfigure[]{
      \scalebox{\twoFeynScale}{
        \includegraphics{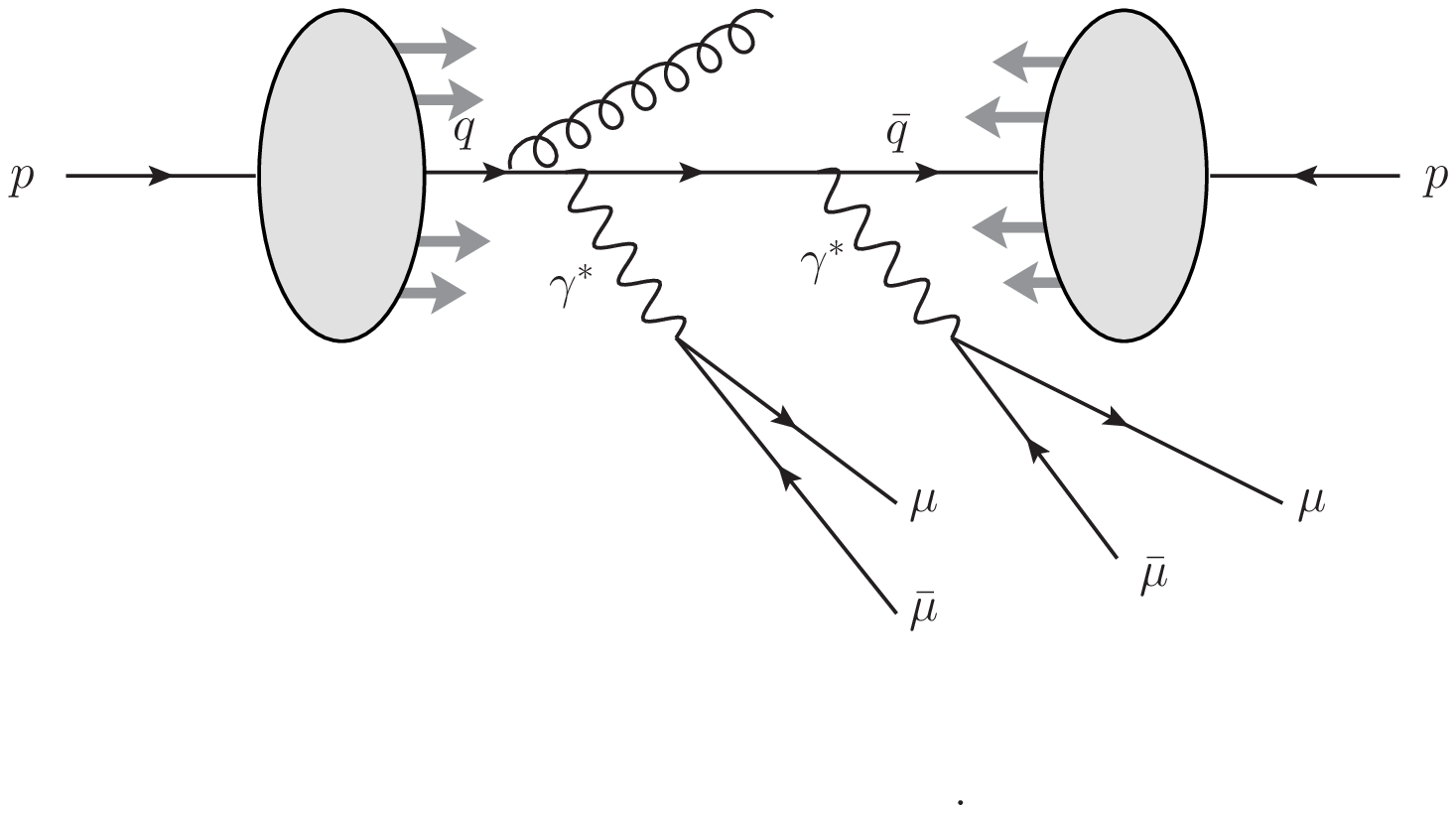}
      }
    }
    \subfigure[]{
      \scalebox{\twoFeynScale}{
        \includegraphics{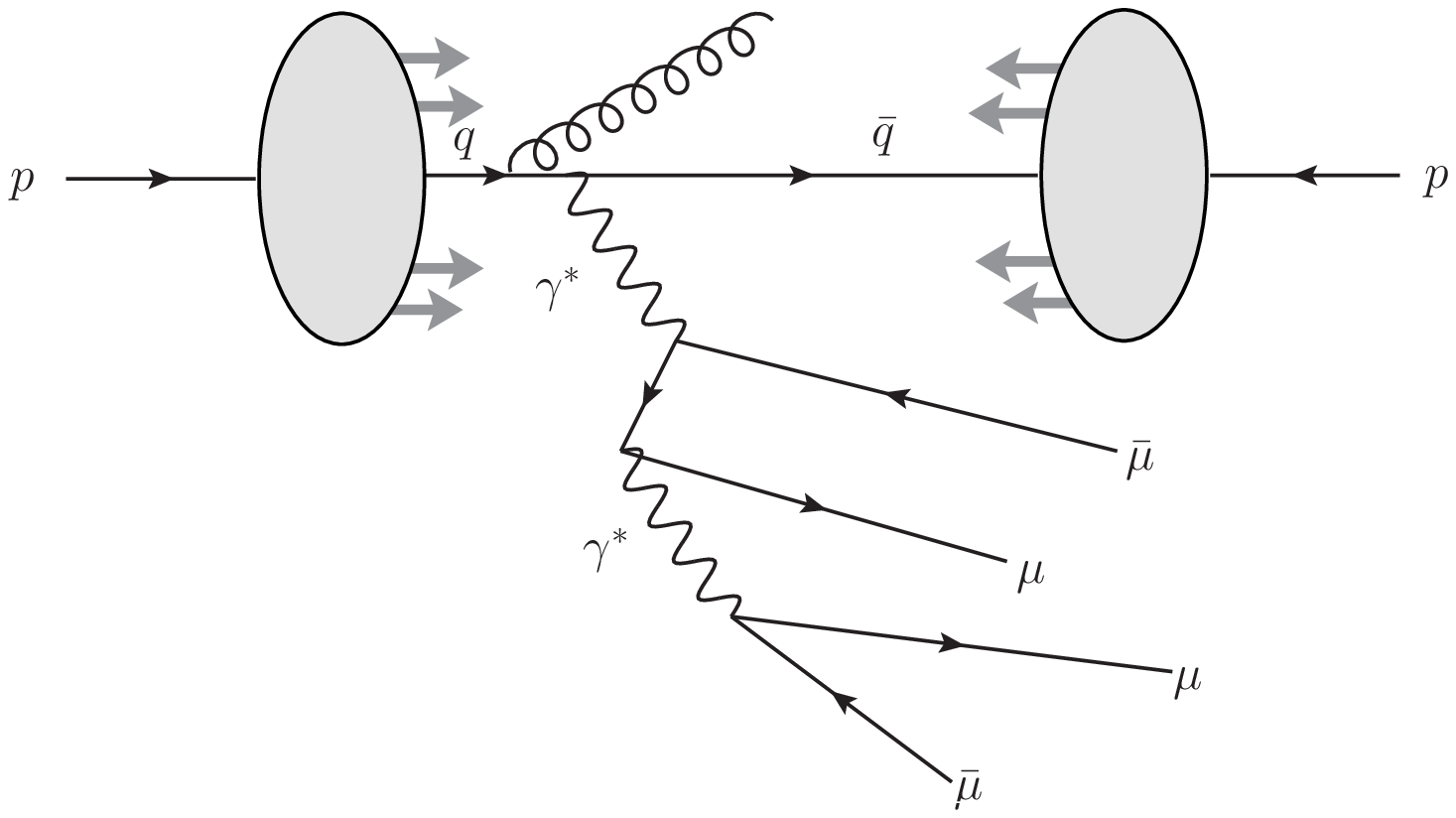}
      }
    }
    \caption{Example Feynman diagrams for the SPS double Drell--Yan
      process.  (a) \Quote{double resonance}--type diagrams, where both
      virtual photons are attached to the initial state quarks, and (b)
      \Quote{single resonance}--type diagrams, where only one virtual
      photon is attached to the quark line.}
    \label{fig:SPSFeynDiag}
  \end{center}
\end{figure*}

Example Feynman diagrams for the SPS production process of two OS muon
pairs are shown in Fig.~\ref{fig:SPSFeynDiag}. They can be classified
into two groups: the \Quote{double resonance}--type diagrams
(Fig.~\ref{fig:SPSFeynDiag}a), where both virtual photons are attached
to the initial state quarks, and the \Quote{single resonance}--type
diagrams (Fig.~\ref{fig:SPSFeynDiag}b), where the quarks fuse into a
$\gams$ before decaying into four muons via an additional $\gams$.  In
our analysis we consider contributions from both types of diagrams, as
opposed to a previous study~\cite{Halzen:1986ue} in which only the
first type of diagrams is included.\footnote{From (electromagnetic)
  charge consideration, it can be seen that these two types of
  diagrams are separately gauge invariant.  In this sense they could
  be considered independently.}

\subsection{Selection of the DPS signal}

As mentioned in the Introduction, in order to observe the DPS signal,
the SPS background has to be low, at least in some region of the phase
space. In the case of two OS muon pair final states, there is no
particular reason why {\it a priori} the SPS production rates should
be suppressed, as opposed to, e.g., the same--sign $W$--pair production,
where due to electromagnetic charge conservation the SPS process can
only occur at higher order in powers of (gauge) couplings than in
DPS. Therefore in studying the final state with two OS muon pairs, it
is important to exploit different kinematical properties of the SPS
and DPS final states.

For the DPS signal, given our independent hard scattering assumption,
each of the two OS muon pairs originating from the two $\gams$ will
exhibit $\pt$ balancing at the parton level.  The presence of initial
state radiation (ISR) will provide a non-zero $\pt$ to each of the
$\gams$.  The independent scattering hypothesis then implies that the
azimuthal angular separation between the two $\gams$s should be
uniformly distributed.  On the contrary, for the SPS background, at
the parton level the transverse momentum is generally only balanced
among all four muons.  An implication is that for the two combinatoric
ways to form two OS muon pairs, which we shall also call $\gams$, both
will lead to a configuration in which the two $\gams$s travel
back--to--back in the transverse plane.  However again this picture
will be distorted in the presence of ISR.  These basic kinematic
differences form the basis for identifying the presence of DPS events
in previous experimental studies.  A number of \Quote{pair wise
  balancing} variables, designed to exploit the different transverse
plane behaviour for the DPS and SPS processes, have been introduced in
the literature.  We shall see that the impact of significant QCD
radiation on the low invariant masses being considered will lead to
significant deviation from the intuitive parton--level picture, making
the usage of such pair wise balancing variables less effective in
discriminating DPS from SPS events.  On the other hand,
\Quote{longitudinal} variables are generally less sensitive to these
effects.  The absolute rapidity difference between two reconstructed
$\jpsi$s was first proposed in \cite{Kom:2011bd} as a method to
extract DPS from SPS events.  We shall see whether a similar strategy
can be used for DDY by constructing a similar variable for the
$\gams$s.

Another complication is that for DDY there is a two--fold
(combinatoric) ambiguity in grouping the four muons into two OS pairs.
This ambiguity is absent for double $\jpsi$, since the \Quote{correct}
pairing must have $\mup\mum$ invariant masses lying within a mass
window close to the physical $\jpsi$ mass.  This leads to a different
set of cuts that must be applied to the DDY, which will result in
different kinematical distributions.  We shall discuss the impact of
these effects in more detail in the following sections.

So far, we have considered the irreducible SPS background only.  In
the low mass region, semi--leptonic decay of heavy quarks and hadron
mis--identification can fake the signal, and are important sources of
background.  These backgrounds have been studied in LHCb in the
context of single DY \cite{LHCb-CONFNote_DY}, and it was found that a
multi--variate selection strategy based on the isolated nature of the
muons produced in DY processes can be used to achieve purities of 95\%
down to $m_{\mup\mum}=2.5$ GeV.  A similar method could presumably be
used for DDY, however a detailed study is beyond the scope of the
present study.


\section{Event simulation}\label{sec:eventSim}

We now discuss how the DPS signal and SPS background events are
generated.  The predictions for the DPS signal are obtained using the
multi--parton scattering model in \herwigv\cite{Bahr:2008pv}, which
allows for generation of two hard scatterings in every event.  We use
the following (default \herwig) $\alpha(Q^2)$ definition:
$\alpha(Q^2)=\alpha(0)/(1-K(Q^2))$, where $\alpha(0)=1/137.04$, and
$K(Q^2)=\frac{\alpha(0)}{3\pi}(13.4955 + 3{\rm log}(Q^2/{\rm GeV}^2))
+ 0.00165 + 0.00299{\rm log}(1.0 + {Q^2}/{\rm GeV}^2)$.  We use
\mstwnl PDFs~\cite{Martin:2009iq}, and set the factorization scale
$\muf$ for both 2--to--2 single DY subprocesses at $\muf=\rtshat$,
which can be different for the two subprocesses in the same event.
\herwig enables us to take higher order QCD effects into account in
the parton shower.  In principle \herwig can also include effects of
intrinsic $\pt$--smearing of incoming partons.  However currently
$\pt$--smearing is applied only to the \Quote{primary} hard process.
In order to provide intrinsic $\pt$--smearings for {\it both} hard
subprocesses, we switch off the default intrinsic Gaussian $\pt$
functionality during event generation, and implement independent $\pt$
smearings to the two hard subprocesses using the same Gaussian model.
For illustration, the root--mean--square intrinsic $\pt$ in this
model, $\sigma$, is set to a conservative value of 2 GeV.

The SPS events with two OS muon pairs are generated using the package
\madgraphv\cite{madgraph}.  Here the same $\alpha(Q^2)$ definition as
in the DPS process is used.  We include both \Quote{double resonance}
and \Quote{single resonance}--type diagrams, as discussed in Section
\ref{sec:process}. In the low invariant mass region being considered
in this analysis, it is sufficient to consider only the Feynman
diagrams with two $\gams$, \ie no diagrams with (virtual) $Z$--bosons
are included.  For consistency, in the experimental analysis, the
contributions from processes involving $Z$'s have to be eliminated.
This can be achieved by imposing an upper cut on the invariant mass of
the four muons, $m_{4\mu}$.  Such a cut is also useful in enhancing
the signal--to--background (S/B) ratio, as the DPS signal is expected
to have comparatively low four--muon invariant mass.  The
factorization scale is set at $\muf=\rtshat$, \ie $m_{4\mu}$.  The
parton level events generated by \madgraph are interfaced to \herwig
for parton showering.  As in the DPS process, the \mstwnl PDFs are
used and an intrinsic Gaussian $\pt$ smearing is applied after event
generation, with the parameter $\sigma$ set to 2 GeV.

We now discuss event selection.  For the muons to be observed in the
LHCb detector, the pseudorapidity ($\eta$) and $\pt$ of each muon has
to obey:
\begin{itemize}
\item $1.9 < \eta < 4.9$\,,
\item $\pt > 1$ GeV\,.
\end{itemize}
Apart from the muons coming from the decay of $\gams$, the simulated
events can also contain additional muons generated by the decay of
hadrons formed in the hadronisation process.  To be conservative, we
select events with exactly four muons passing the above acceptance
cuts, and that they form two OS muon pairs.  Additionally, we
request that the events correspond to processes with $\gams$ decaying
into muon pairs by avoiding the low mass hadronic resonance and the
$Z$ resonance regions with the following cuts:
\begin{itemize}
\item $\mmumu > 4$ GeV for {\it all} four $\dmu$ combinations\,,
\item $ m_{4\mu} < 40$ GeV\,.
\end{itemize}
Note that the cut $\mmumu>4$ GeV must apply to all $\mup\mum$
combinations, instead of only two of them.  This is because there are
potentially dangerous backgrounds involving low mass hadronic
resonances, for example $\jpsi$.  These resonances could lead to high
invariant masses from {\it mismatched} $\mup\mum$ pairs, \ie pairs
that do not come from the decay of the same resonance.  A cut on all
$\mup\mum$ pairs will however also apply to the {\it true} pairing,
which of course has a low $\mup\mum$ invariant mass, and reject the
event as a result.

Furthermore, contributions from $\Upsilon(1S)$ and its higher
resonances $\Upsilon(2S)$ and $\Upsilon(3S)$, when combined with other
source of muons, can lead to two OS muon pairs.  The $\Upsilon$
background can be suppressed by rejecting events with
\begin{itemize}
\item $9.2$ GeV $ < \mmumu < 10.5$ GeV for {\it any} of the four
  $\dmu$ combinations,
\end{itemize}
where the mass window is obtained from the recently published result
by LHCb on $\Upsilon$ production~\cite{LHCb-CONFNote_Epsilon}.

We will refer to the cuts above as the set of \Quote{basic cuts} for the
DDY analysis. We assume 100\% detection efficiency for an event passing the above
cuts.  The cross sections for the production of four--muon final states
in $pp$ collisions at 7 and 14 TeV, after applying basic cuts, are displayed in
Table~\ref{tab:xsec_cuts_DDY}.

\begin{table}[!t]
  \centering
  \begin{tabular}{|c|r@{.}l|r@{.}l|}
    \hline
    \multicolumn{5}{|c|}{DDY cross sections [fb] at LHCb}\\
    \hline
      & \multicolumn{2}{|c|}{\;\;\;\;\;\;\;\;\;DPS\;\;\;\;\;\;\;\;\;} 
    & \multicolumn{2}{|c|}{\;\;\;\;\;\;\;\;\;SPS\;\;\;\;\;\;\;\;\;} \\
    \hline
    7 TeV  & \;\;\;\;\;\;\;\;\;0&08 
           & \;\;\;\;\;\;\;\;\;0&43\\
    14 TeV & \;\;\;\;\;\;\;\;\;0&16 
           & \;\;\;\;\;\;\;\;\;0&68\\
    \hline
  \end{tabular}
  \caption{DPS and SPS DDY cross sections (in $\fb$) for $pp$
    collisions at 7 and 14 TeV.  For clarity, only the results with
    parton shower and intrinsic Gaussian $\pt$ broadening $\sigma=2$
    GeV are shown.}
  \label{tab:xsec_cuts_DDY}
\end{table}

For comparison, we also present predictions for four--muon final
states originating from a decay of a $\jpsi$--pair, produced in $pp$
collisions at 7 and 14 TeV. Our line of analysis follows the one
of~\cite{Kom:2011bd}, where we studied the double $\jpsi$ production
at LHCb for a collision energy of 7 TeV, \ie we apply the same methods
of calculations and the same parameter values as in~\cite{Kom:2011bd}.
As for the DDY analysis, the set of single muon cut is given by:
\begin{itemize}
\item $1.9 < \eta < 4.9$\,,
\item $\pt > 1$ GeV\,.
\end{itemize}
The narrower range $2<\eta< 4.5$ in the analysis of~\cite{Kom:2011bd}
was motivated by the cuts used by the LHCb collaboration in the first
experimental analysis of double $\jpsi$
production~\cite{LHCb-CONFNote_DJpsi}.  Out of the two combinatoric
ways to form two OS pairs, the combination with the two invariant
masses closest to the physical $\jpsi$ masses is chosen. This
requirement is sufficient to discriminate between the combinations, as
the OS muon pairs originate from $\jpsi$ resonances.  In our event
simulation, we do not include effects due to finite detector
resolution. As for the DDY analysis, \mstwnl PDFs are used and 100\%
reconstruction and detection efficiency is assumed. The DPS and SPS
cross sections obtained after basic cuts and the invariant--mass
selection criterion are applied are displayed in
Table~\ref{tab:xsec_cuts_DJpsi}.

\begin{table}[t]
  \centering
  \begin{tabular}{|c|r@{.}l|r@{.}l|}
    \hline
    \multicolumn{5}{|c|}{double $\jpsi$ cross sections [pb] at LHCb}\\
    \hline
    & \multicolumn{2}{|c|}{\;\;\;\;\;\;\;\;\;DPS\;\;\;\;\;\;\;\;\;} 
    & \multicolumn{2}{|c|}{\;\;\;\;\;\;\;\;\;SPS\;\;\;\;\;\;\;\;\;} \\
    \hline
    7 TeV  & \;\;\;\;\;\;\;\;\;3&16 
           & \;\;\;\;\;\;\;\;\;1&70\\
    14 TeV & \;\;\;\;\;\;\;\;\;7&69 
           & \;\;\;\;\;\;\;\;\;2&62\\
    \hline
  \end{tabular}
  \caption{DPS and SPS double $\jpsi$ cross sections (in $\pb$),
    including the branching ratio $\br(\jpsi\to\mup\mum)$.  For
    clarity, only the results with parton shower and intrinsic $\pt$
    broadening $\sigma=2$ GeV are shown.  }
  \label{tab:xsec_cuts_DJpsi}
\end{table}

At present, it is not yet understood how to systematically include
higher order effects within the DPS framework.  The inclusion of the
parton shower in our simulations may be considered as a possible first
step towards predictions with higher order corrections taken into
account.  However, the shower evolution does not change the total
event rate of the parton level, if integrated over the full phase
space.  Therefore it is illustrative to consider the size of the
next--to--leading (NLO) $K$--factors for the total cross sections.

In the context of our calculations, the appropriate K--factor would be
defined by the ratio of the total cross section at NLO calculated using
NLO pdfs to the LO total cross section using NLO pdfs.  For the $pp
\to \gams\gams \to e^+e^-\mu^+\mu^-$ process at 14 TeV, calculations
with the \mcfm package~\cite{Campbell:1999ah} give, for the set of
basic cuts considered here,
$K_{\rm NLO}=1.7$.

For comparison, the K--factor for {\it single} DY, defined in the same
way as above, is $K_{\rm NLO}=3.0$, when the cross sections are calculated
with invariant mass $m_{\mup\mum}$ between 4 and 40 GeV.  This could
indicate an enhanced S/B ratio if higher order effects were taken into
account.

Based on the DPS and SPS rates quoted in Tables
\,\ref{tab:xsec_cuts_DDY} and \,\ref{tab:xsec_cuts_DJpsi}, we see that
the measurement of DPS in four--muon final states will be more
challenging for the DDY process, when compared to double $\jpsi$
production.  In particular, the rates for DPS DDY are of $\ord(0.1)$
$\fb$ and $\ord(0.2)$ $\fb$ at the 7 and 14 TeV LHC respectively, with
relatively low S/B ratios.  However possible improvements could be
made.  For example, it might be possible to better understand the
$\Upsilon$ background by investigating the behaviour of the inclusive
$\Upsilon$ sample \cite{LHCb-CONFNote_Epsilon} that contains an extra
pair of muons when the \Quote{basic cuts} are applied.

We also recall that in our calculation we assume a fixed value for
$\sigeff=14.5$ $\mb$.  This value might be process and scale
dependent, and as discussed in Section \ref{sec:process} is a
parameter to be determined by experiment.  Therefore in the next
section we will focus on studying kinematical distributions and
investigate if further kinematical cuts can be introduced to improve
the S/B ratio, as well as facilitating the extraction of $\sigeff$.
With luminosity of a few inverse femtobarn, the number of DDY events
expected at the 7 TeV LHC is likely to be too low.  For this reason
only results for the 14 TeV LHC will be presented.


\section{Kinematic distributions for the DDY process}\label{sec:kinematics}

In this section we study differential cross sections for the
production of four--muon final states originating from a decay of two
$\gams$ after the set of basic cuts has been applied.  To see the
effects of parton showering and intrinsic $\pt$ smearing, we present
three sets of predictions for the SPS and DPS production mechanism:
\begin{itemize}
\item Parton level only\,(labelled as \Quote{PL} in the plots),
\item Parton level + parton shower\,(labelled \Quote{PS}),
\item Parton level + parton shower + intrinsic $\pt$ smearing\,(labelled \Quote{PS+$\sigma$}).
\end{itemize}

\begin{figure*}[!]
  \begin{center}
    \subfigure[]{
      \scalebox{\twoscale}{
        \includegraphics{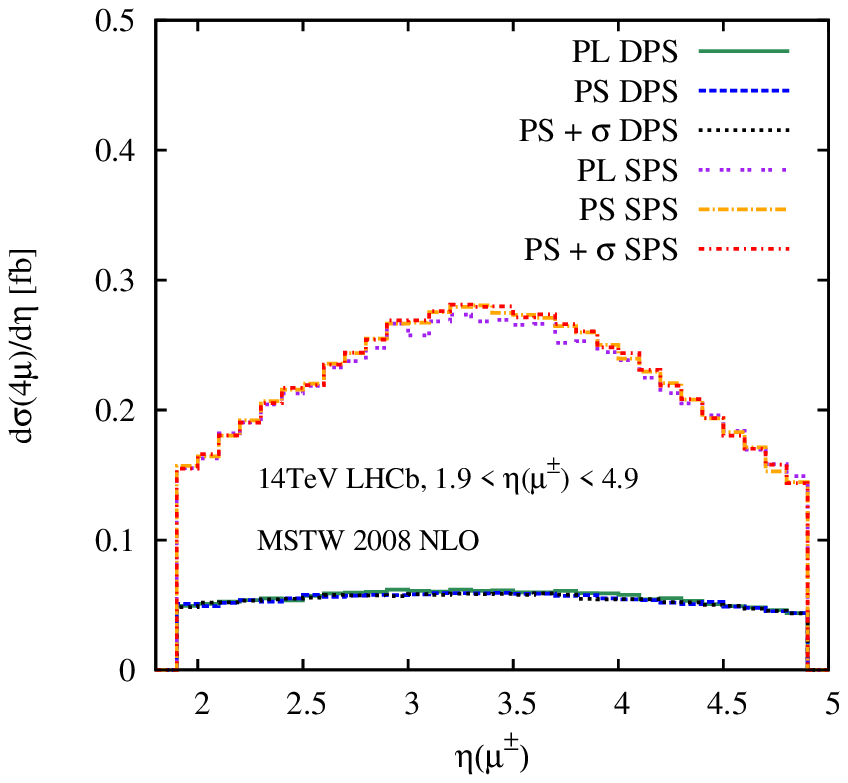}
      }
    }
    \subfigure[]{
      \scalebox{\twoscale}{
        \includegraphics{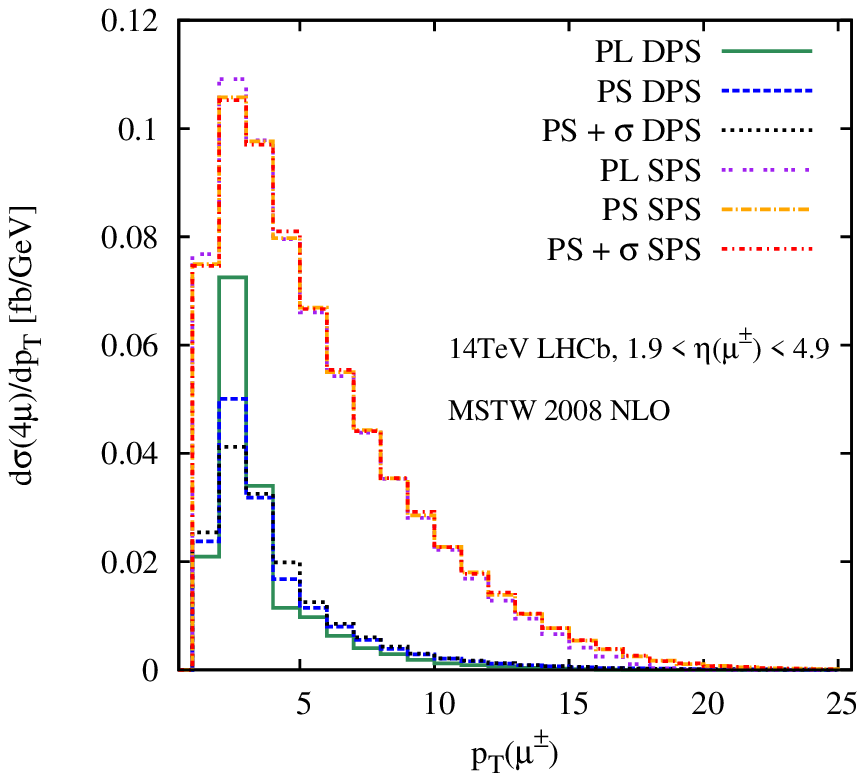}
      }
    }
    \caption{The DPS and SPS single $\mu$ distributions in (a) $\eta$
      and (b) $\pt$ for the process $pp \to \gams \gams \to \mup\mum
      \mup\mum$ at the LHC with 14 TeV.  
    }\label{fig:DDY_singleMu}
  \end{center}
\end{figure*}

The rapidity and transverse momentum distributions for a single $\mu$
are shown in Fig.~\ref{fig:DDY_singleMu}. We see that in the entire
$\eta$ and $\pt$ range considered here, the SPS background dominates
over the DPS signal.  The $\eta$ distribution for the SPS background
is more central within the allowed $\eta$ range.  This is because the
four muons originate from a single hard process, and it is more likely
for all of them to be within the $\eta$ acceptance if the four--muon
system falls in the centre of this allowed region.  The $\pt$
distributions of SPS and DPS muons peak at about the same value, with
the SPS distribution being slightly harder.  Apart from the DPS $\pt$
distribution, we see that the other distributions shown in
Fig.~\ref{fig:DDY_singleMu} are not sensitive to ISR or intrinsic
$\pt$.

Next we study the separation of DPS and SPS events using their
different kinematic behaviour in the transverse plane.  As discussed
in Section \ref{sec:process}, at parton level, provided that the
$\mup\mum$ pairs are grouped correctly, the vector $\pt$ sum of the
two OS pairs in a DPS event will be balanced {\it separately}, \ie
$\vpt=0$.  For the SPS, such $\pt$ balancing is generally achieved
only among all 4 muons, which implies that the $\vpt$ of the two
$\mup\mum$ pairs will be opposite to each other, resulting in a
back--to--back configuration in the transverse plane.  These
differences can in principle be used to distinguish the DPS from the
SPS process.  In addition, these can be used to find the $\mup\mum$
pairing that corresponds to the actual DPS process, and allow further
analysis based on the kinematic properties of these pairs, \ie the
$\gams$ in our present context.

To select one of the two combinatoric ways to form two OS muon
pairs, we select the combination that minimises the following
pair--wise balancing variable ($S$): 
\bea
S&\equiv&\frac{1}{2}\left(\frac{|\vec{\pta}+\vec{\ptb}|}{\pta+\ptb}+\frac{|\vec{\ptc}+\vec{\ptd}|}{\ptc+\ptd}\right)\,,\qquad 0<S<1
\label{eq:Sdef}
\eea
For the DPS signal, at parton level the correct pairing always leads
to $S=0$, while ISR and intrinsic $\pt$ effects will distort the
picture.  Other variables similar to $S$ have been defined in the
literature, for \eg those in Ref.~\cite{D0}.  We have checked that
their performance are similar to the $S$ variable defined in
Eq.~\ref{eq:Sdef}

\begin{figure*}[!t]
  \begin{center}
    \subfigure[]{
      \scalebox{\twoscale}{
        \includegraphics{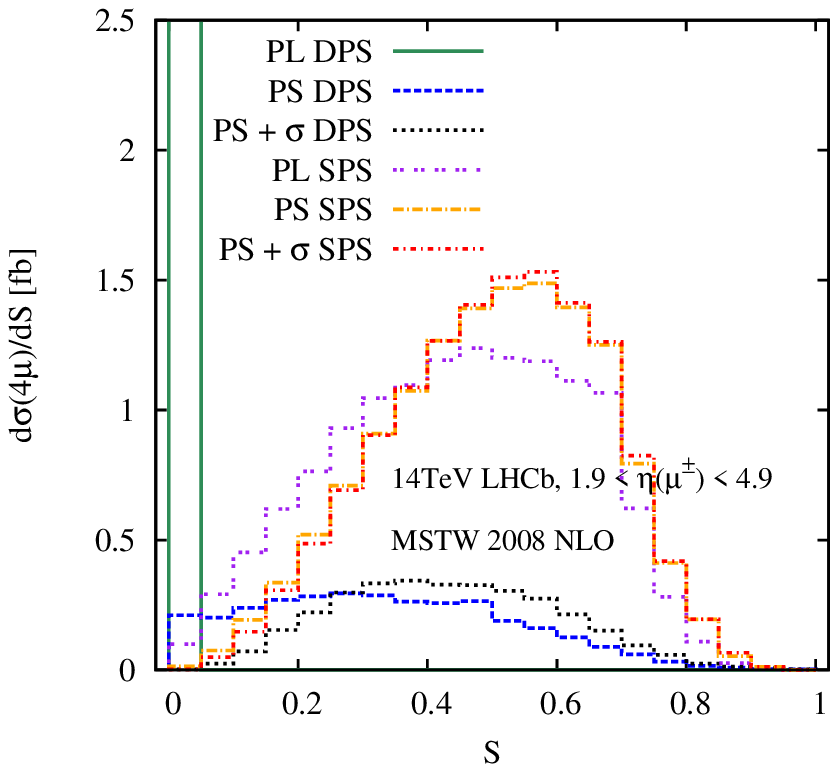}
      }
    }
    \subfigure[]{
      \scalebox{\twoscale}{
        \includegraphics{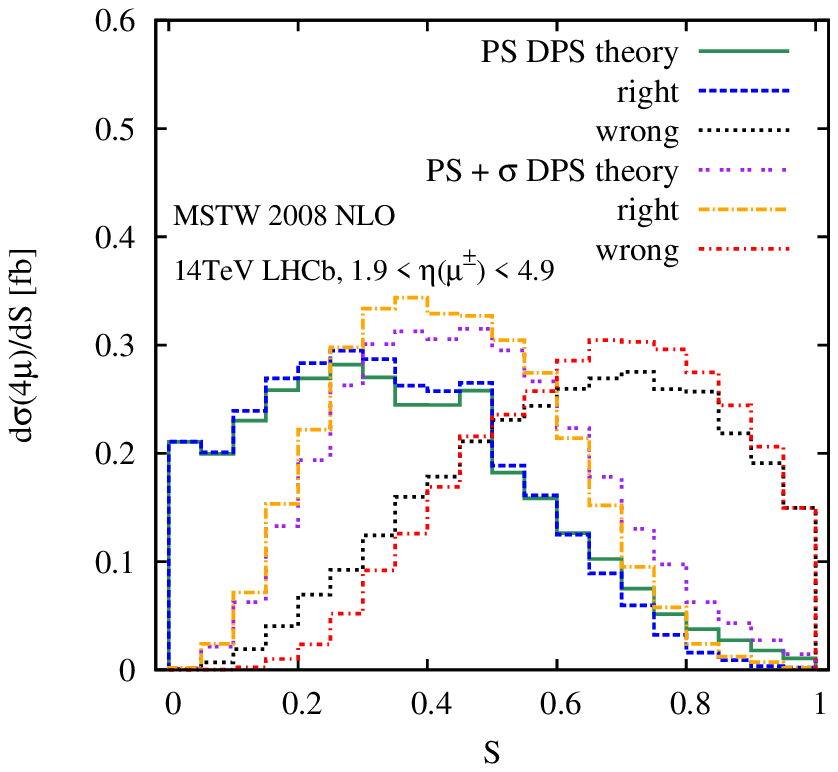}
      }
    }
    \caption{(a) The DPS and SPS differential distributions in the
      $\pt$--balance variable $S$ for the $\mup\mum$ combination that
      minimises $S$, and (b) comparison of the DPS differential
      distributions for the \Quote{right}, obtained by minimizing $S$,
      and \Quote{wrong} $\mup\mum$ pairs with the theoretical
      distribution obtained by always pairing a $\mup\mum$ pair from
      the same $\gams$ together.  In plot (b), the top (bottom) three
      keys correspond to parton shower simulations without (including)
      intrinsic $\pt$ smearing.}\label{fig:DDY_S}
  \end{center}
\end{figure*}

The $S$ distributions are displayed in Fig.~\ref{fig:DDY_S}.  As we
can see, the DPS parton level result is in strict contrast to the SPS
process.  In fact, from Fig.~\ref{fig:DDY_S}a it can be seen that the
values of $S$ close to 0.5--0.7 are the most preferred for SPS,
whereas $S$ peaks sharply in the lowest bin for the DPS at parton
level, as expected.  Unfortunately, radiation and intrinsic $\pt$
effects change the picture dramatically.  While the $S$ distribution
for the SPS process gets modified in a relatively moderate way, the
additional gluon radiation in the DPS process spoils the $\pt$ balance
to a large extent, flattening and spreading the distribution up to
high values of $S$. This, of course, significantly decreases the power
of the variable $S$ to discriminate between the SPS and DPS
mechanisms.  In Fig.~\ref{fig:DDY_S}b we compare the $S$ distributions
between the pairings obtained in the above procedure (\Quote{right}),
the other pairing (\Quote{wrong}) and the theoretical distribution
(\Quote{theory}) obtained by insisting that the $\mup\mum$ pair comes
from the same $\gams$.  We see that the \Quote{theory} and
\Quote{right} distributions are broadly similar, indicating the
effectiveness of the minimum $S$ criterion to pair the OS muons coming
from decay of the same $\gams$ in DPS, despite the presence of
radiation and intrinsic $\pt$ effects. We estimate that using this
requirement leads to selecting the correct $\mup\mum$ pair for around
75\% of the DPS event sample.  For the SPS background, due to much
more complicated kinematical features, the minimum $S$ criterion does
not select any particular physical SPS configurations.

\begin{figure*}[!t]
  \begin{center}
    \scalebox{\scale}{
      \includegraphics{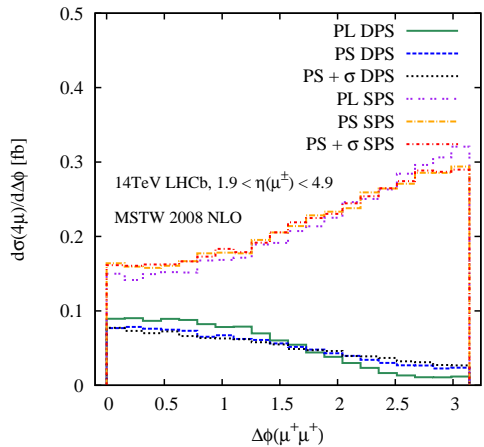}
    }
    \caption{The DPS and SPS differential distributions in the angle
      $\dphi$ between the two $\mup$  for the process $pp \to \gams \gams \to \mup\mum
      \mup\mum$ at the LHC with 14 TeV.}\label{fig:DDY_dPhi_pp}
  \end{center}
\end{figure*}

\begin{figure*}[!t]
  \begin{center}
    \subfigure[]{
      \scalebox{\twoscale}{
        \includegraphics{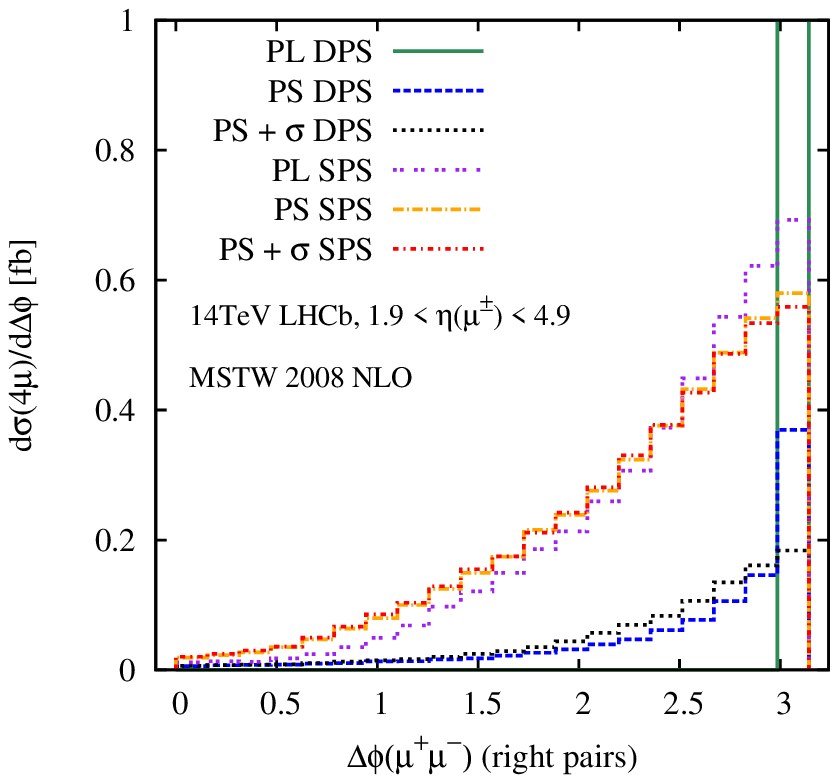}
      }
    }
    \subfigure[]{
      \scalebox{\twoscale}{
        \includegraphics{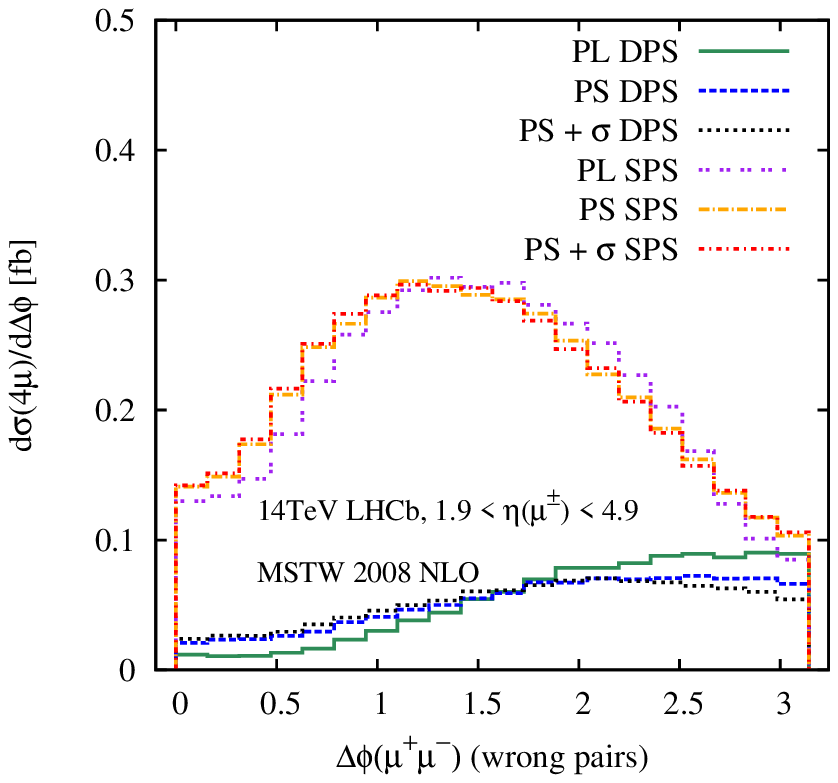}
      }
    }
    \caption{The DPS and SPS differential distributions in the angle
      $\dphi$ between the opposite--sign muons for the (a)
      \Quote{right} and (b) \Quote{wrong} pairs for the process $pp
      \to \gams \gams \to \mup\mum \mup\mum$ at the LHC with 14
      TeV.}\label{fig:DDY_dPhi_pm}
  \end{center}
\end{figure*}

We next study the distribution in the angle on the transverse plane
between the two same--sign muons, $\dphi$, here chosen to be
$\mup\mup$, \cf Fig.~\ref{fig:DDY_dPhi_pp}.  The SPS events tend to
occur with bigger angular separation, which can be traced back to the
tendency for the $\mup\mum$ pairs to be back--to--back in the
transverse plane at parton level.  For the DPS signal, one would
expect that independent scatterings lead a flat $\dphi$ distribution.
However, the requirement that all four $\mup\mum$ pairs pass the
invariant mass cut preferentially selects events with all OS
$\mu$--pairs apart, resulting in a smaller angular separation for
same--sign muons.  The effect for the opposite--sign muons can be
observed in Fig.~\ref{fig:DDY_dPhi_pm}, which shows the angular
distributions for the \Quote{right} and \Quote{wrong} combinations of
$\mup\mum$ pairs. The \Quote{right} pairs minimize $S$, making it more
likely for DDY that the two paired muons come from decay of the same
$\gams$. In particular, at LO where the DY $\gams$ have no $\pt$,
choosing $S$ to be minimal tends to pick out only those $\mup\mum$
combinations which are back--to--back, see Fig.~\ref{fig:DDY_dPhi_pm}a.
The preference towards larger angular separation between \Quote{right}
DPS pairs remains after including radiation and smearing
effects. However, the SPS events also tend to have larger angular
distributions between the OS muons in pairs that minimize the value of
$S$. This tendency is no longer true for SPS \Quote{wrong} $\mup\mum$
pairs, \ie the combination with a larger value of $S$.  In fact, the
qualitative effect is opposite to that of $\mup\mup$ shown in
Fig.~\ref{fig:DDY_dPhi_pp}, which is not unexpected since both
distributions involve azimuthal angular separations between muons from
the wrong pairs, and would be correlated to certain extent.

\begin{figure*}[!h]
  \begin{center}
    \subfigure[]{
      \scalebox{\twoscale}{
        \includegraphics{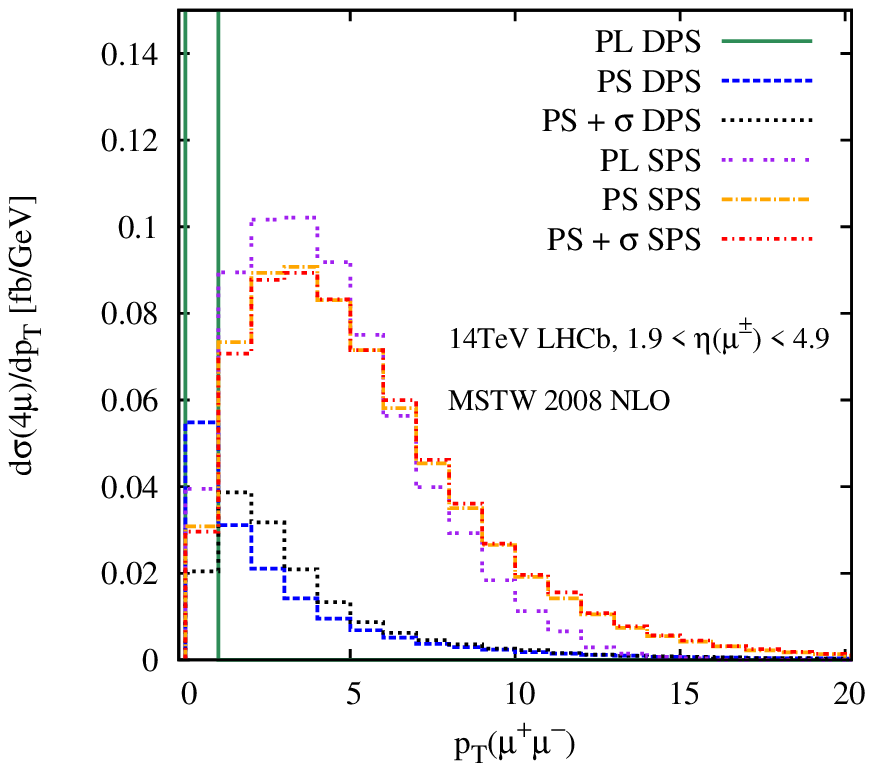}
      }
    }
    \subfigure[]{
      \scalebox{\twoscale}{
        \includegraphics{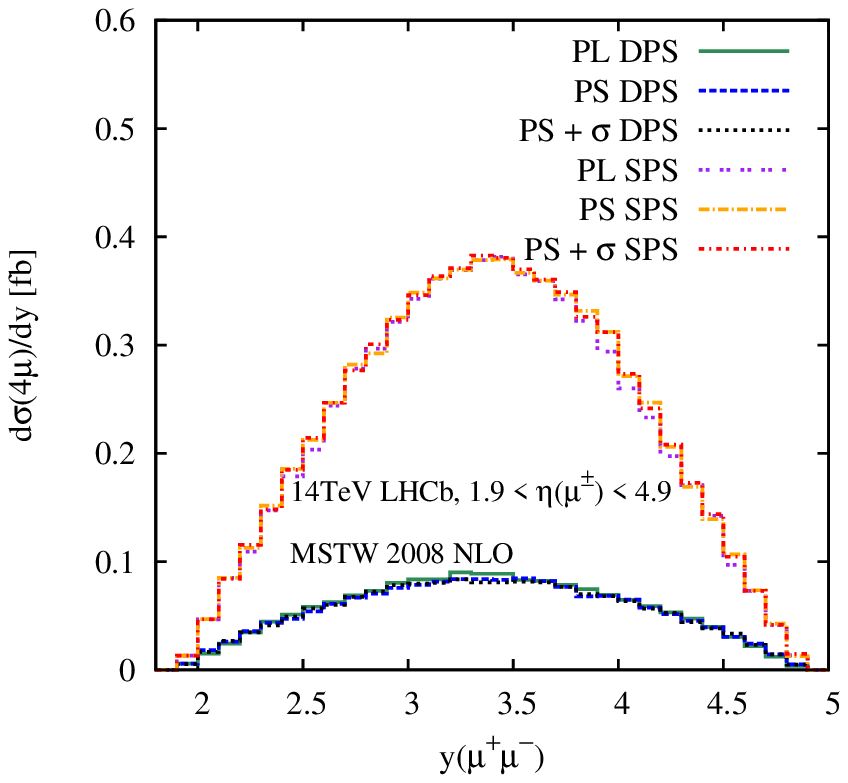}
      }
    }
    \caption{The DPS and SPS differential distributions in (a) $\pt$
      and (b) $y$ for the \Quote{right} $\mup\mum$ pairs for the
      process $pp \to \gams \gams \to \mup\mum \mup\mum$ at the LHC
      with 14 TeV.}\label{fig:DDY_vecPtvecY}
  \end{center}
\end{figure*}

\begin{figure*}[!h]
  \begin{center}
    \subfigure[]{
      \scalebox{\twoscale}{
        \includegraphics{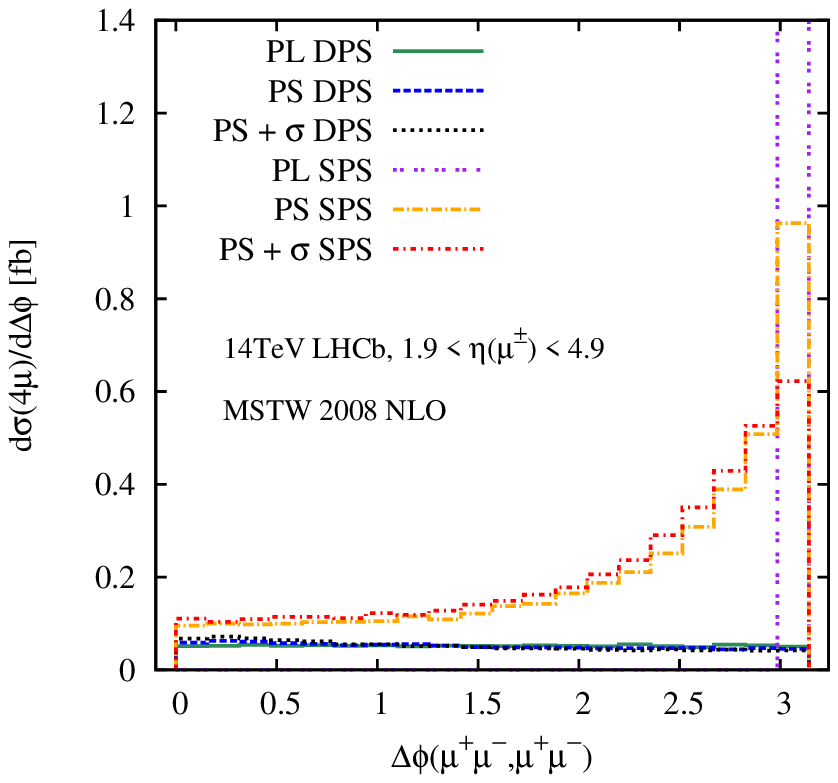}
      }
    }
    \subfigure[]{
      \scalebox{\twoscale}{
        \includegraphics{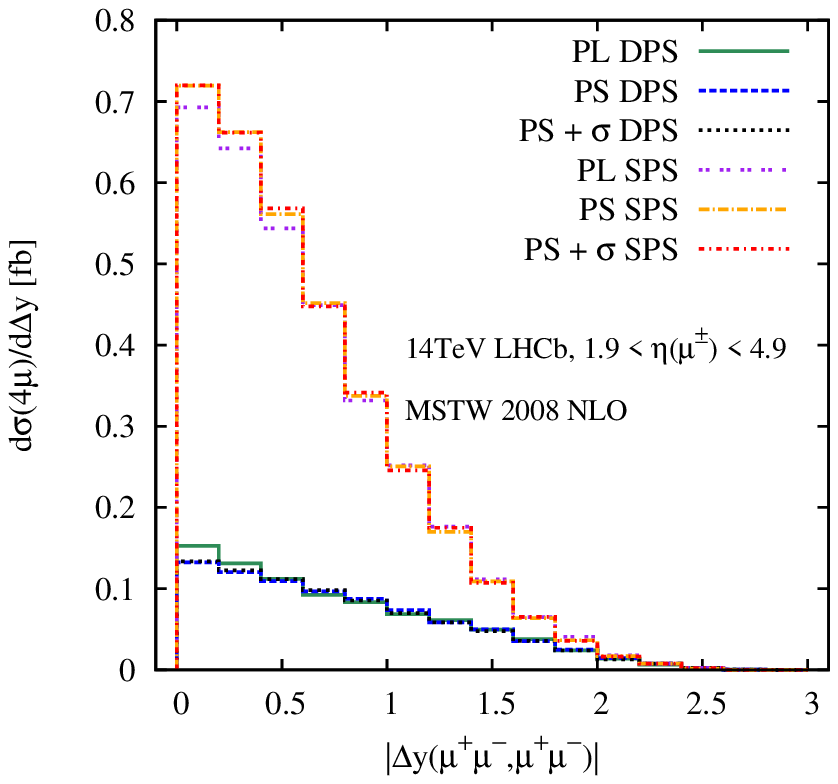}
      }
    }
    \caption{The DPS and SPS differential distributions in (a) the
      angle $\dphi$ between the two $\mup\mum$ pairs and (b) the
      absolute value of the difference in rapidity $|\drap|$ for the
      \Quote{right} muon--pairs for the process $pp \to \gams \gams
      \to \mup\mum \mup\mum$ at the LHC with 14
      TeV.}\label{fig:DDY_vecDPhivecDEta}
  \end{center}
\end{figure*}

Further differential distributions involving $\mup\mum$ pairs,
selected using the $S$--variable criterion, are shown in
Figs.~\ref{fig:DDY_vecPtvecY} ($\pt$ and rapidity $y$) and
~\ref{fig:DDY_vecDPhivecDEta} ($\dphi$ and $\drap$). As expected, the
$\pt$ distribution of the DPS muon--pairs is peaked at very low
values, lower than those for the SPS distribution, even after the
showering and smearing effects are included, \cf
Fig.~\ref{fig:DDY_vecPtvecY}a.  In particular, in the low $\pt$ region
an excess of events compared to a pure SPS model prediction might be
observed.  As shown in Fig.~\ref{fig:DDY_vecPtvecY}b, both the SPS and
DPS have fairly symmetric rapidity distributions for the constructed
$\mup\mum$ pairs, with very similar distributions.  In
Fig.~\ref{fig:DDY_vecDPhivecDEta}a we present the distributions for
the angle (in the transverse plane) between the $\mup\mum$ pairs from
the \Quote{right} combination, $\dphi (\mup\mum,\mup\mum)$. The DPS
distribution is relatively uniform over the whole range of $\dphi
(\mup\mum,\mup\mum)$, consistent with the assumption of two
independent scatterings in the DDY process. The SPS contribution, on
the other hand, peaks at $\dphi (\mup\mum,\mup\mum)=\pi$, which is a
consequence of the transverse momentum conservation for the SPS
process at leading order, as already discussed.
Fig.~\ref{fig:DDY_vecDPhivecDEta}b shows the properties of the SPS and
DPS processes with respect to correlations in the longitudinal
direction, or more precisely the absolute value of the difference
between the rapidities of the two $\mup\mum$ pairs, $|\drap
(\mup\mum,\mup\mum)|$.  As opposed to distributions depending on
variables measured in the transverse plane, distributions involving
rapidity variables exhibit very little dependence on radiation and
intrinsic smearing, an effect first noticed and put to use
in~\cite{Kom:2011bd}.  We see that as $|\drap|$ increases, the SPS
contribution falls more steeply than that of DPS, although the
difference is not as strong as in the double $\jpsi$ analysis
in~\cite{Kom:2011bd}.

\begin{figure*}[!t]
  \begin{center}
    \subfigure[]{
      \scalebox{\twoscale}{
        \includegraphics{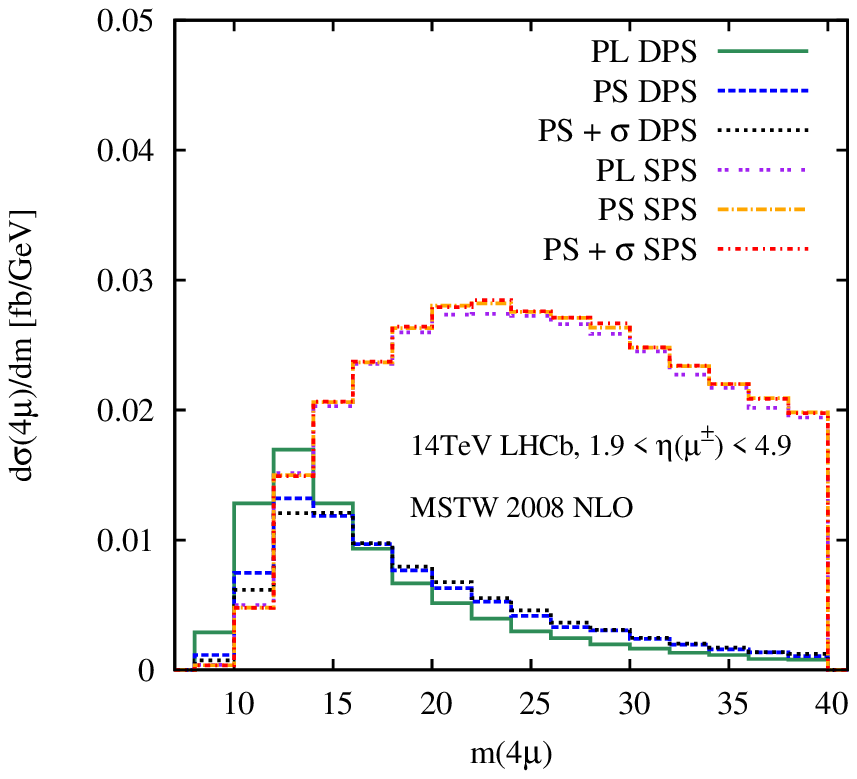}
      }
    }
    \subfigure[]{
      \scalebox{\twoscale}{
        \includegraphics{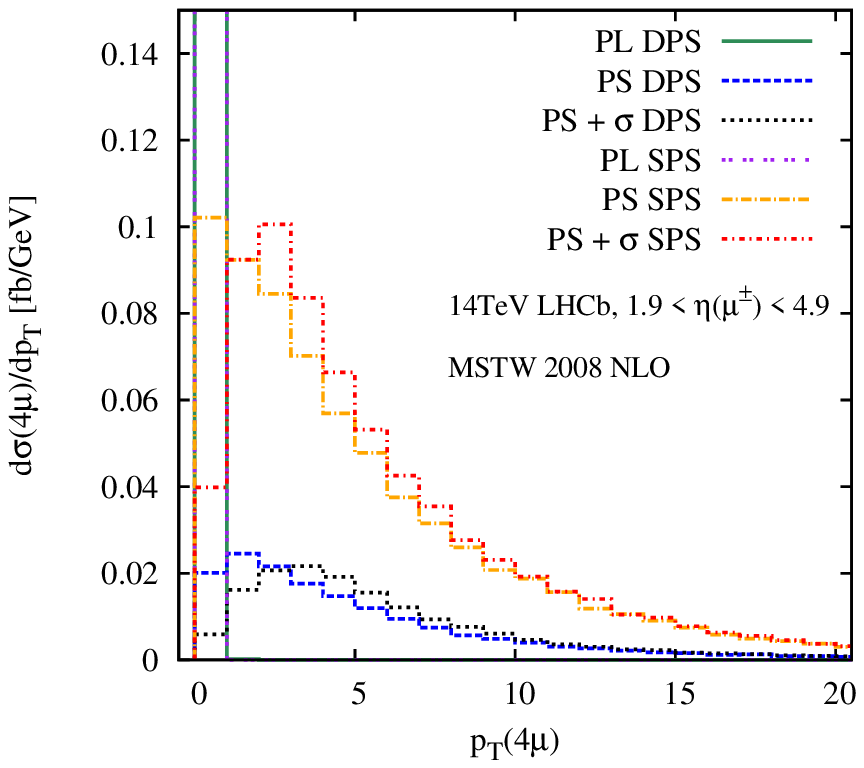}
      }
    }
    \caption{The DPS and SPS differential distributions for all four
      muons in (a) the invariant mass $m(4\mu)$ and (b) the transverse
      momentum $\pt(4\mu)$ for the process $pp \to \gams \gams \to
      \mup\mum \mup\mum$ at the LHC with 14 TeV.}\label{fig:DDY_4Mu}
  \end{center}
\end{figure*}

Yet another possibility for studying the DPS and SPS distributions is
offered by the distributions sensitive to quantities constructed from
four vectors of all muons in the final state. In
Fig.~\ref{fig:DDY_4Mu} we investigate the invariant mass $m(4\mu)$ and
the transverse momentum $\pt (4\mu)$ distributions of all four
muons. A significant difference in the invariant mass distribution can
be observed for the SPS and DPS mechanisms, \cf
Fig.~\ref{fig:DDY_4Mu}a, and an excess of DPS events might be observed
in the low $m(4\mu)$ region.  This can be traced back to the lower cut
on the $\mup\mum$ invariant mass, which, as already noted, favours
configurations with all OS $\mu$--pairs apart. For DPS the impact of
the cut is relatively mild, as the independent scattering hypothesis
implies that OS pairs from different hard scatterings tend to be well
separated.  This is not the case for the SPS background, which results
in higher invariant mass of all leptons in the final state.  From this
figure, it can be inferred that events with $m(4\mu)>40$ GeV are
dominated by the SPS background, and hence could form a control region
for the SPS background estimation.  A more aggressive cut on max
$m(4\mu)$, down to around 30 GeV, appears to be possible.  Such a cut
would reduce the SPS background by around 30\%, while having a
relatively mild impact on the signal.  However, we remain conservative
here and refrain from doing so.  The $\pt (4\mu)$ distribution shown
in Fig.~\ref{fig:DDY_4Mu}b measures the sensitivity of the SPS and DPS
to extra radiation, simulated by parton showering, and intrinsic
smearing. Both distributions are affected by these effects in a
similar way, with the peak of the showered distribution at
approximately the same value and moving to higher $\pt$ as the
smearing is switched on.

\begin{figure*}[!t]
  \begin{center}
    \subfigure[]{
      \scalebox{\twoscale}{
        \includegraphics{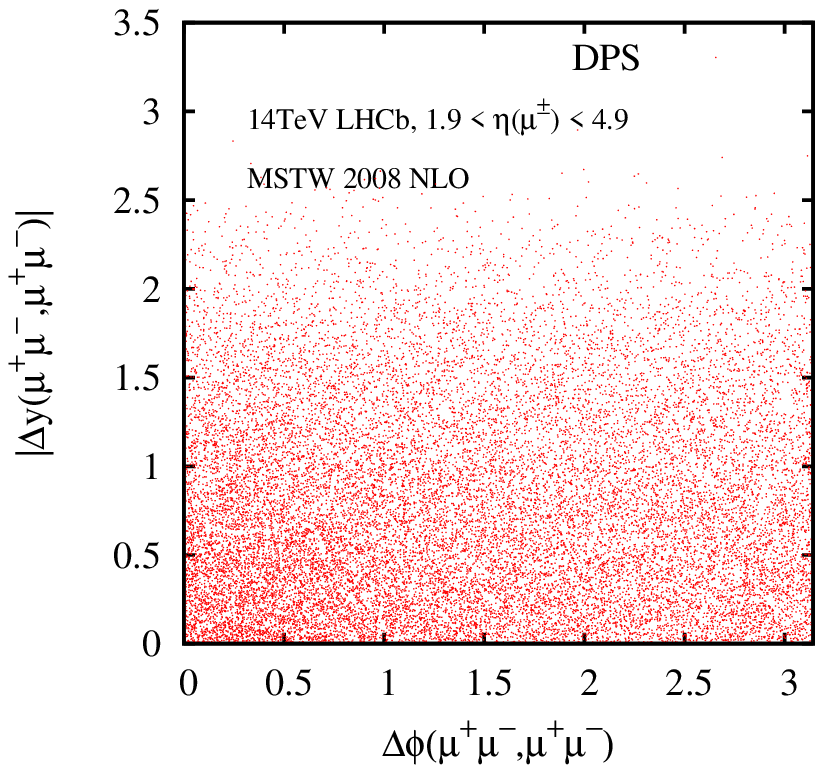}
      }
    }
    \subfigure[]{
      \scalebox{\twoscale}{
        \includegraphics{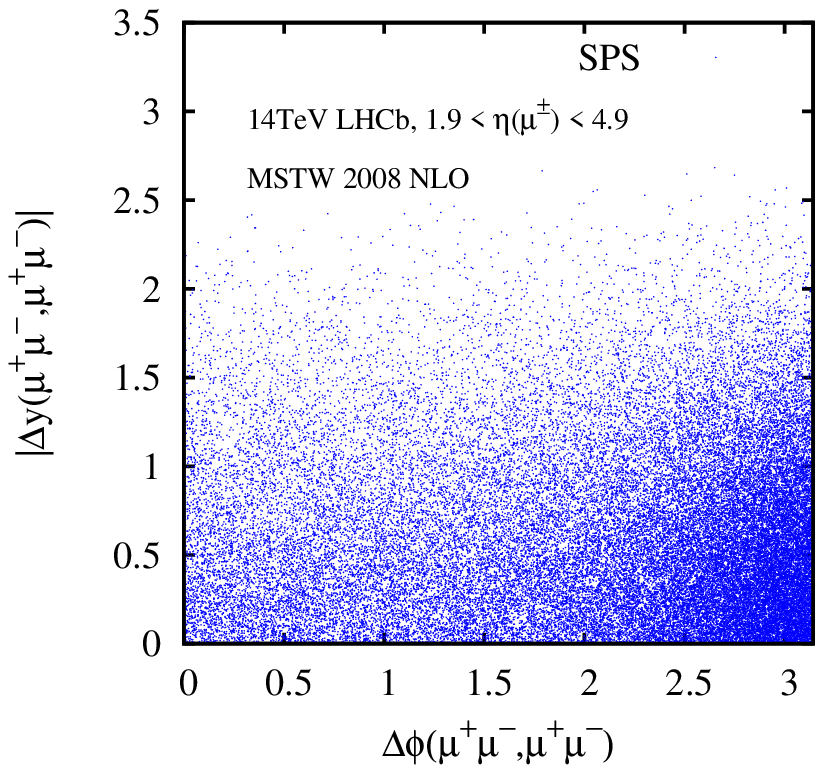}
      }
    }
    \caption{Scatter plots on the ($\dphi (\mup\mum,\mup\mum) $, $
      |\drap (\mup\mum,\mup\mum)|$) plane for the (a) DPS and (b) SPS
      mechanism for the process $pp \to \gams \gams \to \mup\mum
      \mup\mum$ at the LHC with 14 TeV.  }\label{fig:DDY_scat}
  \end{center}
\end{figure*}

Since the two types of correlations, \ie in the transverse plane and
in the longitudinal directions, between the muons in the final state
probe different kind of information on the same process, it is
interesting to study the double differential distributions in $\dphi
(\mup\mum,\mup\mum)$ and $|\drap (\mup\mum,\mup\mum)|$. In
Fig.~\ref{fig:DDY_scat} we present scatter plots in the $\dphi
(\mup\mum,\mup\mum)$ and $|\drap (\mup\mum,\mup\mum)|$ plane for the
DPS and SPS samples.  The former appears to be slightly more populated
in the region with small $\dphi$ and $|\drap|$, whereas events of the
latter tend to occupy the corner of large $\dphi (\mup\mum,\mup\mum)$
and small $|\drap(\mup\mum,\mup\mum)|$.

Concluding, observing DPS through four--muon final states from DDY
scatterings seems to be a challenging task due to its low production
rate, \cf Table~\ref{tab:xsec_cuts_DDY}.  Although an excess of DPS
events could be observed in the low $\pt (\mup\mum)$ and $m(4\mu)$
regions, the studied differential distributions between the SPS and
DPS event samples do not show kinematical differences that are
striking enough to justify putting additional hard cuts (perhaps with
the exception of $m(4\mu)$) at the cost of further reducing the signal
yield.

On the other hand, we see that in addition to $\pt (\mup\mum)$ and
$m(4\mu)$, variables such as $S$, $\dphi(\mup\mum,\mup\mum)$,
$|\drap(\mup\mum,\mup\mum)|$ or $\dphi (\mup\mum)$ of the wrong
muon--pairs also show moderate, non--trivial differences between the
DPS and SPS components, and we are able to understand the origin of
their different distributions.  To fully utilize these differences, it
would be more effective to apply the template method, as is done for
example in the $\gamma + 3$j studies in CDF and D0 to extract the DPS
fraction.  Moreover, we see from the double differential distribution
in Fig.~\ref{fig:DDY_scat} that additional information can be
obtained by simultaneously considering more than one kinematic
quantity.


\section{Kinematic distributions for double $\jpsi$ production}\label{sec:kinematics_jpsi}

As seen in Section \ref{sec:eventSim}, after applying sets of basic
cuts the total cross sections for double $\jpsi$ production at LHCb
are significantly higher than the equivalent cross sections for the
DDY process.  Even more importantly the S/B ratio is also better.
This makes double $\jpsi$ production an excellent candidate process to
observe DPS~\cite{Kom:2011bd, Baranov:2011ch, Novoselov:2011ff}.

Methods to study double $\jpsi$ and to extract the DPS signal from the
SPS background have been discussed in some detail
in~\cite{Kom:2011bd}.  Here we extend this study by comparing the
different properties between double $\jpsi$ and DDY in $pp$ collisions
at 14 TeV.

There are a few important differences between double $\jpsi$ and DDY.
First, their production is initiated by different partons: while
double $\jpsi$ production is driven by the gluon--gluon fusion
process, both SPS DDY and the single DY subprocess in DPS DDY require,
at lowest order, a quark and an antiquark.  The longitudinal momentum
densities of these partons are different, which would lead to
different rapidity distributions for the $\mup\mum$ pairs.  Different
initial--state partons also affect the parton shower, and hence $\pt$.
However, the hard scales of the double $\jpsi$ and DDY are also
different, which will in turn affect the amount of QCD radiation.  The
combined effect has an impact on $\pt$ as well as azimuthal angular
separations in the transverse plane of the reconstructed
objects.\footnote{In the present study we do not attempt to
  disentangle the effects on $\pt$ from these sources.}

\begin{figure*}[!ht]
  \begin{center}
    \subfigure[]{
      \scalebox{\twoscale}{
        \includegraphics{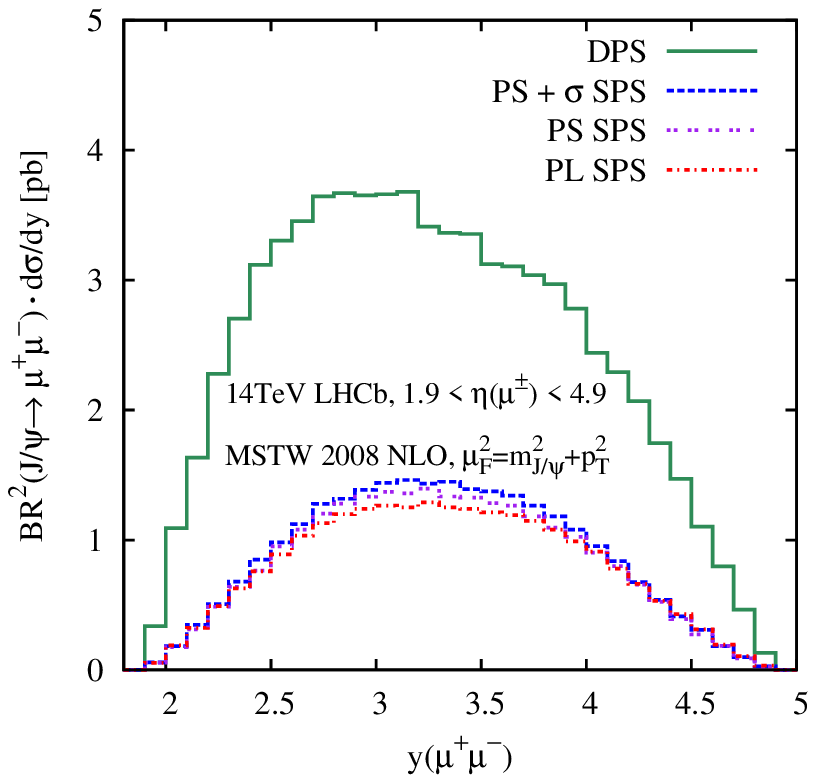}
      }
    }
    \subfigure[]{
      \scalebox{\twoscale}{
        \includegraphics{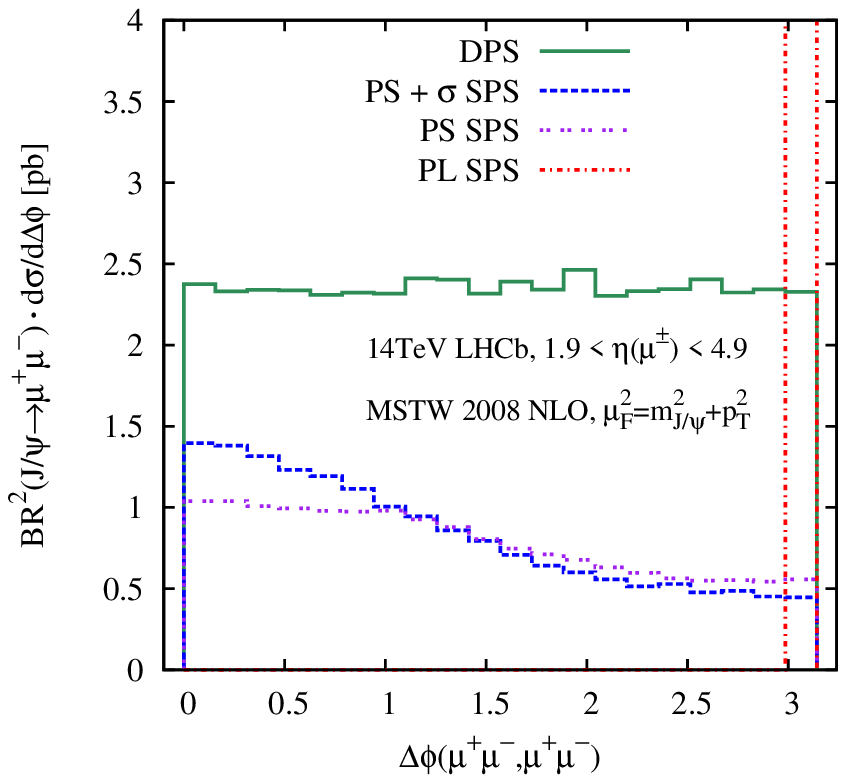}
      }
    }
    \caption{The DPS and SPS differential distributions in (a) $y$ and
      (b) $\dphi(\mup\mum,\mup\mum)$ of $\mup\mum$ pairs for the
      process $pp \to \jpsi \jpsi \to \mup\mum \mup\mum$ at the LHC
      with 14 TeV.}\label{fig:DJpsi_vecPtvecY}
  \end{center}
\end{figure*}

In Fig.~\ref{fig:DJpsi_vecPtvecY} we show the rapidity and
$\dphi(\mup\mum,\mup\mum)$ distributions of the $\mup\mum$ pairs from
double $\jpsi$ production.  Here the invariant mass of a $\mup\mum$
pair readily determines if they originate from a $\jpsi$ resonance or
not, and hence there is no ambiguity in pairing up the OS muons.
Compared with DDY (\cf Fig.~\ref{fig:DDY_vecPtvecY}b), we see in
Fig.~\ref{fig:DJpsi_vecPtvecY}a that both the double $\jpsi$ SPS and
DPS are more central, with the effect of SPS stronger.  This is due to
the smaller gluon densities at high parton $x$ compared to the light
(anti) quarks, making high $\eta$ values for the $\mup\mum$ pairs less
likely.  In Fig.~\ref{fig:DJpsi_vecPtvecY}b the impact of QCD
radiation on $\dphi(\mup\mum,\mup\mum)$ is clearly seen, where the
distribution including parton shower peaks slightly towards zero,
instead of $\pi$, the parton level result.  This should be compared
with the corresponding figure for DDY,
Fig.~\ref{fig:DDY_vecDPhivecDEta}a, which shows a milder radiation
effect.

A consequence of the well defined way to pair up the OS muons is that
it is not necessary to introduce the variable $S$ as a way to group
the muons, although it could still be used to help distinguish the SPS
and DPS components.  Since for the SPS DDY process the pairing
according to value of $S$ does not correspond to finding two $\mu$'s
from a decay of the same $\gams$, more significant differences between
DDY and double $\jpsi$ are expected for the SPS background.

\begin{figure*}[!ht]
  \begin{center}
    \subfigure[]{
      \scalebox{\twoscale}{
        \includegraphics{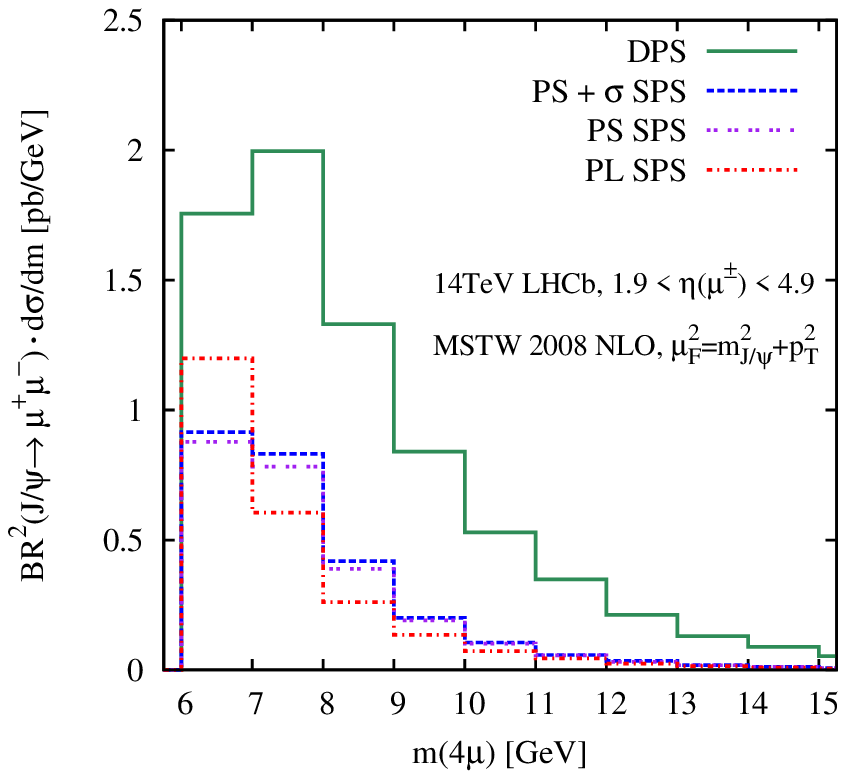}
      }
    }
    \subfigure[]{
      \scalebox{\twoscale}{
        \includegraphics{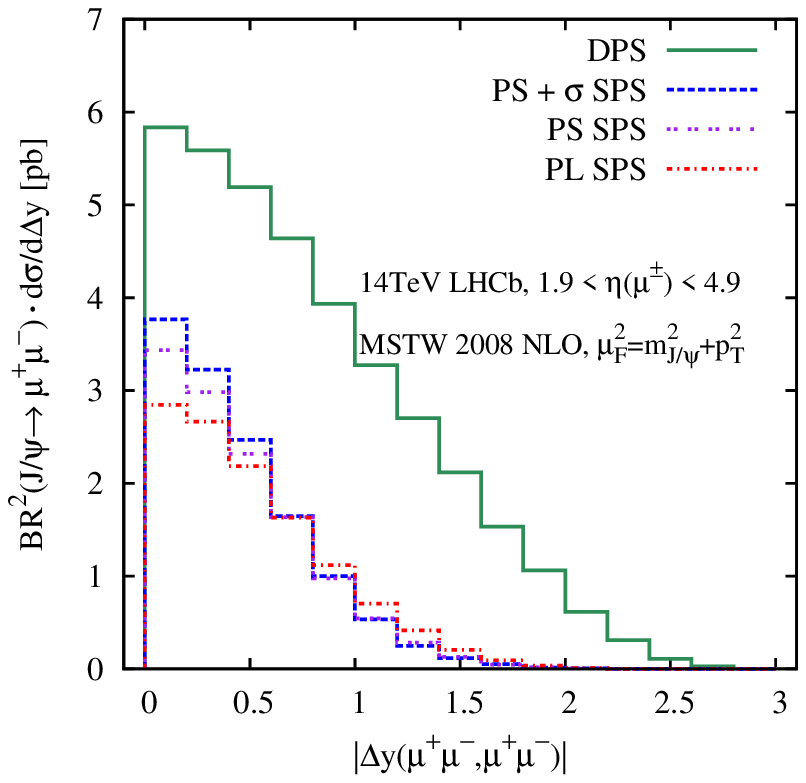}
      }
    }
    \caption{The DPS and SPS differential distributions for (a) the
      invariant mass $m(4\mu)$ and (b) the absolute value of the
      difference in rapidity $|\drap(\mup\mum,\mup\mum)|$ for the
      $\mup\mum$ pairs for the process $pp \to \jpsi \jpsi \to
      \mup\mum \mup\mum$ at the LHC with 14 TeV.
    }\label{fig:DJpsi_4Mu}
  \end{center}
\end{figure*}

The four--muon invariant mass distribution for double $\jpsi$
production is shown in Fig.~\ref{fig:DJpsi_4Mu}a.  The DPS component
peaks at higher value of $m(4\mu)$ than the SPS component.  The
reverse is observed for DDY, \cf Fig.~\ref{fig:DDY_4Mu}a.  This can
partly be attributed to the $\mup\mum$ invariant mass cut applied in
the DDY analysis, which will tend to push the SPS $m(4\mu)$
distribution towards higher values.  Of course there is no equivalent
mass cut in the $\jpsi$ study, since there is no ambiguity in pairing
up the muons.  A quantity related to $m(4\mu)$ is
$|\drap(\mup\mum,\mup\mum)|$, as in the centre of mass frame, a larger
$m(4\mu)$ will implies a larger $|\drap(\mup\mum,\mup\mum)|$ given a
fixed $\pt(\mup\mum)$.  This distribution is displayed in
Fig.~\ref{fig:DJpsi_4Mu}b.  The observation for double $\jpsi$ of a
more steeply falling $|\drap|$ for the SPS is consistent with the more
steeply falling $m(4\mu)$.  For DDY, the broader SPS $m(4\mu)$
distribution implies that $|\drap|$ would be more slowly falling, thus
making $|\drap|$ less effective as a variable to extract DPS from SPS
component in DDY.

Finally, we note that overall our current studies reinforce the
statements made in~\cite{Kom:2011bd} regarding the importance of
radiation and intrinsic $\pt$ smearing effects for the observables
involving quantities measured in the transverse plane.  At the same
time, we confirm the relative insensitivity to these effects on the
variables measuring correlations in the longitudinal direction.
Therefore it could be beneficial to investigate in more detail how
\Quote{longitudinal variables} may be used to facilitate other DPS
searches in the future.


\section{Conclusions and outlook}\label{sec:conclusion}

Measurements undertaken at the LHC will be crucial for improving the
theoretical description of multiple parton scattering.  In this paper
we have studied the production of two OS muon pairs via two hard
scatterings of the DY type.  DY production is considered a standard
candle process at hadron colliders. The leptonic final states provide
very clean experimental signatures, while the status of theoretical
predictions is under control. All this makes the production of two OS
muon--pairs a model candidate process for probing the nature of DPS.

Depending on the invariant mass of the OS muon pairs, they may
originate either from continuum ($\gams$) or a hadronic resonance, \eg
$\jpsi$.  In the course of our study, we have found that at 7 TeV, the
DDY cross section and luminosity are too low for this process to be
observed with the LHCb detector.  The prospects for observing DDY
should improve at 14 TeV LHC given the increases in cross section, S/B
ratio and luminosity.  On the other hand, the production of the same
final state through a double $\jpsi$ process possesses characteristics
which should allow a measurement of DPS at the LHCb in $pp$ collisions
at both 7 and 14 TeV.  However, we stress that these two processes are
sensitive to different initial state partons ((anti)quark--(anti)quark
for DDY and gluon--gluon for double $\jpsi$).  Hence they probe
different correlation effects and provide complementary input to
double parton distributions.

For the same reason, it might prove valuable to additionally
investigate four muon final states where one muon pair comes from a
decay of $\gams$ while the second originates from the decay of a
low--mass resonance, \eg $\jpsi$.  The cross section for such a
process might be expected to lie between double $\jpsi$ and DDY, and
hence is likely to be observable.  Furthermore, although in this paper
we have concentrated on pairs of $\jpsi$'s as resonant sources of
four--muon final states at low invariant masses, another promising
opportunity to study DPS can be offered by studying higher mass
resonances, in particular involving $\Upsilon$ production.  For
example, measurement of $\sigeff$ for both double $\jpsi$ and double
$\Upsilon$ processes could provide information on the scale dependence
of the two--gluon parton distribution function.  This would complement
well the study of scale dependence that could be performed in \eg
$\gamma+3$j, $\Wpm\Wpm$ and other jet--based signatures proposed for
the general purpose detectors ATLAS and CMS \cite{MPIatLHC}.

Upon completion of this paper, a suggestion to study double $\Upsilon$
has been put forward in~\cite{Novoselov:2011ff}.


\section*{Acknowledgements}
This work has been supported in part by the Helmholtz Alliance
\Quote{Physics at the Terascale}, the Isaac Newton Trust and the STFC.
AK would like to thank the High Energy Physics Group at the Cavendish
Laboratory for hospitality.  CHK thanks the Institute for Theoretical
Particle Physics and Cosmology at RWTH Aachen University and the
Particle and Astroparticle Physics group at MPIK Heidelberg for
hospitality while part of the work was carried out.


\end{document}
